# Effective Use of Human Computer Interaction in Digital Academic Supportive Devices

[1]Thuseethan, S., [2]Kuhanesan, S.

[1]Department of Computing and Information Systems, Sabaragamuwa University of Sri Lanka
Belihuloya, Sri Lanka.
thuseethan@gmail.com

[2]Department of Physical Science, Vavuniya Campus of the University of Jaffna
Vavuniya, Sri Lanka.
kuhan9@yahoo.com

**Abstract:** *In this research, a literature in human-computer interaction is reviewed and the technology aspect of human computer interaction related with digital academic supportive devices is also analyzed. According to all these concerns, recommendations to design good human-computer digital academic supportive devices are analyzed and proposed. Due to improvements in both hardware and software, digital devices have unveiled continuous advances in efficiency and processing capacity. However, many of these systems are also becoming larger and increasingly more complex. Although such complexity usually poses no difficulties for many users, it often creates barriers for academic users while using digital devices. Usually, in designing those digital devices, the human-computer interaction is left behind without consideration. To achieve dependable, usable, and well-engineered interactive digital academic supportive devices requires applied human computer interaction research and awareness of its issues.*

## 1. Introduction

Today, computer and information technology has a significant role in education through utilizing e-learning environments and different computer based academic supportive devices. The involvement of Information Communication Technology leads education environments towards an era of electronic academic supportive devices.

The digital devices such as notebook, tablet pcs and handheld portable devices such as smartphones have become almost usual equipment in higher education (Weaver and Nilson, 2005). The usage of electronic academic devices in learning environment is significant, because it offers attractive, more realistic and interesting teaching facility. At the same time usage of digital devices in the classroom is intended to enhance the learning environment for all students. It was also evident that the use of digital devices in classroom was effective in enhancing motivation, the ability to apply course based understanding, and whole academic achievement amongst students (Vibert and Mackinnon, 2002).

We roughly categorized the academic digital devices into two broad categories, the devices support the learning process and the devices support teaching process. Devices use in learning process supports the students to gather and expand their knowledge in class rooms while teachers define tasks for students to work. Yuen, Cheung & Tsang (2012) stated that there is a modern interest in using e-textbook to replace paper-based textbook amongst students (Yuen, Cheung & Tsang, 2012). Other type of devices sit between the students and academics in teaching process helps academics in teach students. Wang, Shen, Novak and Pan (2009) stated that digital devices can be used for instant communications among teachers and students.

Even though there are serious preventable problems in those academic supportive devices. Any use of such devices that degrades the efficiency of learning environment, promotes dishonesty or dissatisfaction in the learning process. Normally this kind of degrades happen because of bad design of digital devices, it results with the loss of teaching time. To overrun these problems a need for research for improving the human computer interactions emerges. According to Diaper (2005) the chronology of HCI starts in 1959 with Shakel's paper on "The ergonomics of a computer" which was the first time that these issues were ever addressed. So, for the effective use of academic supportive devices, it should be designed with efficient human computer interactions standards.

The main contribution of this paper is investigation of advantages and disadvantages of the interaction styles in academic supportive devices and the recommendations for designing such devices with the help of good human-computer interaction.

## 2. Literature review

Human-computer interaction can be viewed as two powerful information processors (human and computer) attempting to communicate with each other via a narrow-bandwidth, highly constrained interface (Tufte, 1989). Human-Computer Interaction (HCI) is defined by (ACM SIGCHI, 1996) as "a discipline concerned with the design, evaluation, and implementation of computing systems for human use and with the study of major phenomena surrounding them" Dix et al (1998). As by the definition HCI knows as intersection of different disciplines such as computer science, behavioral science and several others. As the result there is real confusion in what HCI is, a science, a design science or an engineering discipline. Newell & Card (1985) defined HCI as a science; HCI is tempered by approximation, providing engineering-style theories and tools for designers. Carroll & Campbell (1989) defined HCI as a design science, developing a craft-based approach and new research methods to evaluate existing systems in their intended and tasks context, using the results to inform designers for the next generation of systems. The design and strategy of humans and computers intermingling to accomplish work effectively, exposed as an engineering discipline (Long & Dowell, 1989).

Preece(1994) defined as, Human-computer interaction (HCI) is "the discipline of designing, evaluating and implementing interactive computer systems for human use, as well the study of major phenomena surrounding this discipline" (Preece, 1994). As the whole human–computer interaction studies related with both human and machine in combination, it draws from supporting knowledge on both the machine and the human side. Dix(1998) stated that HCI involves the design implementation and evaluation of interactive systems in the context of the users' task and work. Human Computer Interaction basically concerned with the interfaces between man and machine. HCI differs from human factors (or ergonomics) in some ways. HCI mainly focus more on user's perspective, working specifically with computers. HCI also focuses on the implementation mechanisms in software and hardware production to support effective human computer interaction.

While designing devices, the cognitive processes whereby users interact with computers should be considered as main issue because commonly users' attributes do not match to the capabilities of such devices. At the same time such devices may have non-cognitive effects on the user such as users' reaction to virtual worlds. But in most cases human strongly recommend the usual cognitive effects. Reeves & Nass (1996) proved as humans have a robust tendency to react to computers in similar ways as they do to other individuals. By considering the communication between human, interpreting the blend of audio and visual signals holds vital role in understanding communication.

The primary goal of Human Computer Interaction is to improve the interactions between users and computers. It makes computers more operational and receptive to the user's wants. Human computer interaction develops or improves certain goals in designing devices. Five important goals are:
- Safety
- Utility
- Effectiveness
- Efficiency
- Usability

During 1990's the term usability has become popular in all activities in human computer interaction. Diaper stated that the study of HCI became the study of Usability.

### 3.1 Models
A model describes the way of interaction between user and computer.

### 3.1.1 Norman's model of interaction

Norman concentrates on user's view. With the help of psychology, Norman describes the user's cognitive process as the interaction with technology in daily life. Norman's model is divided into two phases: execution and evaluation. Each phase is divided into several steps. As the whole it contains seven distinct steps.
The identified steps are:

- Forming the goal
- Forming the intention
- Specifying an action
- Executing the action
- Perceiving the state of the world
- Interpreting the state of the world
- Evaluating the outcome

### 3.1.2 The Interaction Model

Abowd and Beale defined this framework of interaction as translation between languages. They state both a common interaction framework and a translation within the framework. Abowd and Beale framework concentrate on four components and each has its own unique language. Those are;
- User
- Input
- System
- Output

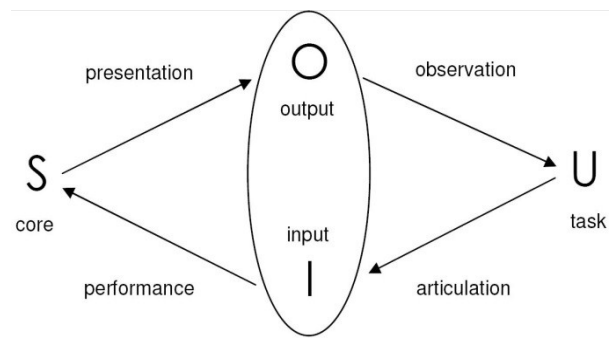

**Figure 1:** Interaction Model: Abowd and Beale Framework

### 3.2 Structure of HCI

HCI, as the name suggests, comprises three major parts within the framework: the user, the computer, and the interaction, indicates the ways they work together to achieve goals. Figure 2 shows three main components of human computer interaction.

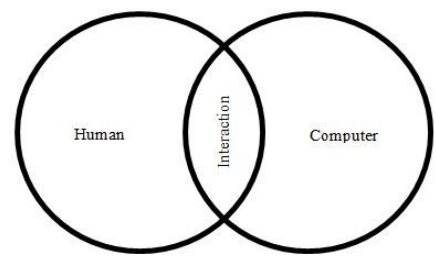

**Figure 2:** Three components of HCI

### 3.2.1 The user

The user analysis is a critical part of user-centered systems design. The public or the user of HCI could be considered as the user of systems. They may vary based on the purposes and task they have in the system. The distinct characterization of users depends on above task and purpose with their experience on it. Danino (2001) stated that the user of HCI is whoever using technology to try to get the job done.

### 3.2.2 The computer

Danino (2001) stated that the computer in HCI denotes to any technology that comprises from desktop computers to generalized computer systems; even an embedded system or an information processing engine can be viewed as "computer". A computer is a device used for general purpose and it carry out several arithmetic and logical operation with the human help. The way of interaction with computers is not limited with traditional shape of the computer because of the incredible technological development. But Human computer Interaction is focused on interfaces involved in man and machine. Each and every device consist some kind of user interface for its usage. Normally it involves with huge amount of interaction.

### 3.2.3 The interaction

The major component in Human Computer Interaction is interaction between man and machine. Normally human interact with other human through speech. At the same time they support their expression with some body gestures, emotions and certain expressions. The non-cognitive properties of a computer system on the user must be looked carefully, because humans always have a solid tendency to respond on a computer in same ways as they react to the practical world (Reeves & Nass, 1996).

## 3. Research on academic supportive devices

### 3.3 Research on devices support the learning process

Input efficiency takes major role in learning supportive devices. In most cases learning supportive devices use to gather or acquire lecture notes in real time. Interaction styles mention to the dissimilar ways of communication in between human and computer. Different systems use different interactions styles. But some common interaction styles are there, those are individually evaluated.

### 3.3.1 Command line languages

This is one popular mode of interaction between humans and computers. Here the computer accepts some typed meaningful commands. Usually user can type one command at a time, thus it is very slow in taking data in. Particular application process or execute the sub sequent inputs given by user and give some feedbacks.
It has some considerable advantages, but the interaction becomes a dialogue only, particularly the human is the lively side and face more workload than computer. Two important pros and cons of command line languages related with academic supportive devices are listed in Table 1.

**Table 1:** Pros and Cons of Command line languages

| Pros | Cons |
|---|---|
| Cheap | Low visibility |
| Flexible | Error handling |

Because of low visibility of command line languages are hard to use in real time environments as well as in academic supportive devices too. Error correction mechanism is very important in academic supportive devices because of its real time usage. But this facility is very much lack in such command line languages.

### 3.3.2 Menus

As the name indicate the menu interface exactly borrows its name from the list of dishes or food items that can be chose in a restaurant or food corner. In same way, a menu interface offers the user with a pre-defined static list of selections in an onscreen fashion. A collection of choices displayed on the screen where the selection and execution of one or more of the selections results in a transformation in the state of the interface (Preece, 1994). There are four brave categories of menus:
- Pull-down menus
- Pop-up menus
- Hierarchical menus
- Contextual menus

Two important pros and one cons of menus related with academic supportive devices are listed in Table 2.

**Table 2:** Pros and Cons of Menus

| Pros | Cons |
|---|---|
| No need to recall | Limited |
| Logical Group | |

Here in menus a very big drawback is it is limited, normally academic inputs are not limited to defined choices. Thus it is not suitable at all for academic supportive devices.

### 3.3.3 Graphical and direct manipulation

The direct manipulations involve in representing the data or information through graphical format. Table 3 indicates the pros and cons of direct manipulation related with learning academic supportive devices.

**Table 3:** Pros and Cons of direct manipulation

| Pros | Cons |
|---|---|
| User sensitive | Limited |
| Flexible | |

Here in direct manipulation a very big drawback is it is limited as like menus, normally academic inputs are not limited to defined choices. Even though it is user sensitive and can easily understand in real time situations.

### 3.3.4 Form fill-in, Question and answer and function keys

By the nature form fill-in, question and answer and function keys are not suitable in academic supportive devices. These three styles of interaction are fully concentrated on a pre-defined flow. But academic supportive devices require a dynamic input flow, it acquire input data in a real time academic environment.

### 3.3.5 Natural language

Natural language processing (NLP) is concerned with human languages such as local languages. It is a field of computer science correlated in the area of Human Computer Interaction.

In learning academic supportive devices, the usage of natural language processing is very much important with comparing other interaction styles. Here we considered the natural language interfaces, a type of interface that allows users to use their own language to input data. Interaction becomes easier in this type of interfaces while using learning academic supportive devices.

### 3.4 Research on devices support the teaching process

By considering teaching supportive devices, output efficiency takes major role rather than input efficiency. Normally these devices use to convoy or spread thoughts of teachers to the learners. In most cases teachers like to have user friendly remote devices or controllers for each teaching supportive devices they use.

Teachers prefer interoperation in between the devices, need to transfer or convert material from one teaching supportive device to another. At the same time they prefer a way of moving materials to learning supportive devices, it enable them to distribute their materials in real time.

Further they pointed out the following functionalities to support their teaching;
- Better graphics resolution
- Widespread and distinguishable buttons or navigations
- Better visibility of text, image, audio and video
- On time graphical outputs
- Convenience and mobility
- Security and safety
- Speech and handwriting recognition

## 4. Review

We found some significant review factors in both learning and teaching supportive devices. Shniderman (1986) stated that the researchers have found that re-design of the human computer interface can create a considerable difference in learning time, performance, speed, error rates and user satisfaction.

In this research we are concerned in understanding suitable strategies for academic supportive device implementations. Some important refined methods of implementing academic supportive devices are given.

### 3.5 Parallel inputs

Devices accept more than one input at a single point of time; each input has been filtered by filters. Finally gathered different input data combine by the combiner, where noises get removed. Before store the data device prompt with feedback to user. Figure 3 indicates the stages in gathering parallel inputs.

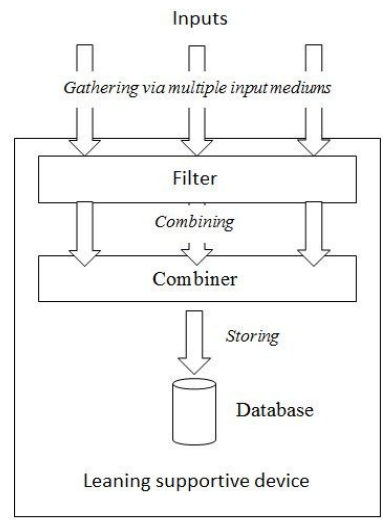

**Figure 3:** Stages: Accepting parallel inputs

### 3.6 Voice recording

Voice is natural way of interaction in academic environments. But continuous voice output is tough to gather or achieve. Even though it is easy to record the voice through interfaces in academic supportive devices with minimum error rate without interruption, much of the argument under voice as input.

Research in finding the way to gather voice input and integrate it into multimode interface is particularly significant. In this case use microphone is simple to get voice input, may have to face problems when having noisy environment. In such cases it is important to integrate parallel input mechanism to avoid loss of data or lecture inputs.

### 3.7 Hand writing recognition

It is also a natural way interaction, even better than voice input. Student can avoid the unwanted conversation here by using handwriting recognition interfaces.

The interfaces with hand writing recognition can be cooperative in reduce the use of other input devices such as mouse and keyboard, and hence reduces the time in inputting. It is useful in solving or writing mathematical or diagrammatical inputs.

### 3.8 Bluetooth connectivity

To increase the interoperability between the devices, the use of Bluetooth is helpful to establish connections. There are three types of connectivity is required in academic environments. Those are;
- One-to-one device connectivity
- One-to-many device connectivity
- Many-to-many device connectivity

In such cases we have to concentrate on security and privacy factors. To have user friendly connectivity it is important to have two different modes of connections such as automatic and manual. Automatic connections allow two or more devices to establish connections without permission, useful when mutual understanding exist already within the users of such devices. On the other hand manual connection could require some authentication of users, useful when teachers allow students to access materials. Connectivity allows academic users the interoperability as well as better co-ordination within that device network.

### 5. Conclusions

In the above research, human computer interaction literature is reviewed as well as technological matters like interaction styles are studied and pros and cons are dogged. And we searched for better interaction styles among the existing ones. At the same time we found dome best "fit" in between a human and a computer in terms of interaction.

While designing moral, effective and user-friendly interfaces for an academic supportive device, several disputes have to be considered. This research suggests a theoretical support in the area of human computer interfacing in designing academic supportive devices. In this paper we have deliberated the promising use of Human Computer Interaction in academic supportive devices to attain top levels of interaction between user and academic devices.

We conclude that to design a worthy human computer interaction, we have to appropriately elect the suitable style of interaction, kind of interface to adequate with the class or group of users it is intended whereas the human issues must be taken into account (Fetaji, M., at al., 2007). Therefore we recommend some important modes of interaction as efficient for academic supportive devices such as parallel input, voice recognition, interoperability among devices and hand writing recognition. We recommend related human-computer interaction design to similar solutions related to academic supportive device designs.

Clearly, we now analyzed all existing techniques in human-computer interaction, in order to increase the efficiency of academic supportive devices. However, the implementation of suggested interaction styles and models offer a sound basis for the future research.

**Author Profile**

**Selvarajah Thuseethan** received the B.Sc. (hons) degree in Information and Communication Technology from Vavuniya Campus of the University of Jaffna in 2013. Now he is working as a Lecturer in computer Science, Department of Computing and Information Systems, Sabaragamuwa University of Sri Lanka.

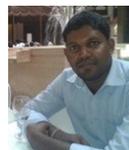

**Sinnathamby Kuhanesan** received the B.Sc. (hons) and M.Phil. degrees in Physics from University of Peradeniya in 1996 and 2005, respectively. Now he is working as a Senior Lecturer in Vavuniya Campus of the University of Jaffna and holds the position Dean, Faculty of Applied Science, Vavuniya Campus of the University of Jaffna.

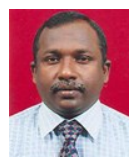